\documentclass{article}

\begin{document}

\title{{\small (SUBMITTED TO GRAVITATION AND COSMOLOGY)}\\
\bigskip
MAGNETIC FIELDS OF ACTIVE GALAXY NUCLEI \\ AND COSMOLOGICAL
MODELS}

\author{Yu.N. Gnedin, T.M. Natsvlishvili, M.Yu. Piotrovich\\
{\small \it Central Astronomical Observatory at Pulkovo,
Saint-Petersburg, Russia}}

\maketitle

\begin{abstract}
We present the review of various methods of detection of magnetic
field strengths in the nearest regions of the active galaxy nuclei
(AGN) which are the high energetic machines. Original
spectropolarimetric method developed in the Pulkovo Observatory is
based on the effect of polarization plane rotation on the mean
free path respect to electron scattering in the plasma. In a
result the spectrum of polarized radiation depends essentially on
radiation frequency unlike the classical case of Thomson
scattering. This fact allows us to determine the magnitude and
geometry of the magnetic field in the region of the optical and
more hard electromagnetic radiation. The results of theoretical
calculations are compared to the results of spectropolarimetric
observations of AGN. The extrapolation of estimated magnitudes of
magnetic fields to the nearest region of supermassive black holes
(SMBH) allows us to determine the magnetic fields of SMBHs. We
developed the method of determining magnetic fields through the
spectrum synchrotron radiation in the region of synchrotron
self-absorption. As the magnitude of magnetic field of the
extragalactic source depends very strongly on the angular size of
extragalactic source and therefore on the photometric distance the
calculated magnetic field magnitudes depends very strongly on the
definite cosmological model. This result allows us to present the
new method for determination of the most important cosmological
parameters including dark matter and dark energy parameters.
\end{abstract}

\section{Introduction}

Magnetic fields are responsible for many aspects of plasma inflow
and outflow in active galactic nuclei (AGNs). Magnetic fields
provide outward angular momentum transport in the accretion
process via magnetorotational instability, with accompanied
dissipative heating that allows further mass-inflow. Strong
poloidal magnetic fields are thought to thread centrifugal winds,
which could carry away angular momentum, while confining and
collimating large-scale outflows and jets seen in many AGNs. The
existence of broad-line region in AGNs may be provided namely by
magnetic pressure. Fig.1 shows a remarkable variety of structure
of typical AGN nearest region including accretion disk, disk
corona, wind stream lines, etc.

First part of our paper is including the short description of new
method of determination of magnetic fields developed in the
Pulkovo Observatory. The main idea of this method is taking into
account the Faraday rotation of the polarization plane in the mean
free path of electron (Thomson) scattering. In a result a
nontrivial wavelength dependence of polarization arises when the
Faraday rotation angle $\psi$ at the Thomson depth $\tau$ (see
Gnedin and Silant'ev 1997)

\begin{equation}
\psi = 0.4 \left(\frac{\lambda}{1 \mu m}\right)^2 \left(\frac{B}{1
G}\right)\tau \cos{\theta}
\label{eq1}
\end{equation}

\noindent is sufficiently large. Here, $\lambda$ is the radiation
wavelength and $\theta$ is the angle between the line of sight and
magnetic field B. We present the results of calculations of
polarized radiation of the accretion disk and accretion outflow
near a supermassive black hole which is thought to be in the
centre of AGNs.

The text of this paper is also including the brief review of
results of polarimetric observations of AGNs and QSOs. The basic
new our idea is to use the future precise measurements of magnetic
fields of AGNs and QSOs for investigating of various cosmological
models and especially for determination of the equation of state
(EOS) of dark energy. The main method for solving this problem is
determination of magnetic field of extragalactic radio structures
using their synchrotron spectra in the region of synchrotron
self-absorption. In this case the magnetic field magnitude depends
very strongly on the value of photometric distance. The last one
depends in this turn on the real cosmological model. It appears
that the magnetic field strength of the given source can be
different in many times for various cosmological models. Thus we
are obtaining the new method that allows us to make the decisive
choice between numerous cosmological models. We expect that this
method could have more advantage over the modern classical SNe and
WMAP methods.

\section{The basic results of polarimetric observations of AGNs}

Fig.2 presents the results of spectropolarimetric observations of
the sample of extragalactic targets chosen by Cabanac et al.
(2005) from Veron-Cetti and Veron (2000) and the Sloan Digital Sky
Survey. The targets included the bright extragalactic sources
having higher probability to show stronger polarization: BAL
quasars and red quasars. The selection criteria of the sample were
detailed by Hutsemekers et al. 2001,2005. To avoid possible
contaminations from the interstellar medium of the Galaxy only the
objects at galactic latitude $b_{gal} > 30$ grad were selected.
Above this galactic latitude the interstellar polarization is
smaller. This sample spans the redshift range $0 < z < 3$
homogeneously. Fig.2 shows the dependence of the quasar sample
polarization on redshift. Overall the quasar polarization degrees
do not correlate with redshift though the slight enhancement of
highly polarized quasars at lower redshift seems to be real. Fig.2
(right panel) presents the histograms of the linear polarization
degree of the quasar sample (black line) and of the star sample
selected over the same region (grey line) are superimposed. Both
panels are used from Cabanac et al. (2005). The star polarization
histogram shows much smaller polarization degrees to that which
corresponds to minor contamination from the interstellar
polarization. Decreasing the number of polarized quasars with
increasing polarization degree may be interpreted as the effect of
depolarization due to magnetic Faraday effect.

Hutsemekers and Lamy, 2001, and Hutsemekers et al. 2005 discovered
coherent orientations of quasar polarization vectors on
cosmological scales. It appears that quasar polarization vectors
are not randomly oriented over the sky with a probability often
excess  of 99.9\%. The polarization vectors appear coherently
oriented or aligned over huge ($\sim 1$ Gpc) regions of the sky
located at both low ($z\sim 0.5$) and high ($z\sim 1.5$) redshifts
and characterized by different preferred directions of the quasar
polarization (Fig.3). The left panel shows a map of the
polarization vectors. The right panel shows the logarithmic
significance level of the coherent orientation for the sample of
213 polarized quasars. Hutsemekers et al.(2005) claimed existence
a regular alternance along the line of sight of regions of
randomly and aligned polarization vectors with typical commoving
length scale of 1.5 Gpc. Furthermore, the mean polarization angle
seems to rotate with redshift at the rate of $\sim 30^{\circ}$ per
Gpc. They excluded the effect of contamination by interstellar
polarization in our Galaxy. Though it is not possible to exclude
the intrinsic quasar origin of this alignment, the other
possibility is that both the polarization vector alignments and
the rotation of the mean polarization angles are due to a physical
mechanism which affects the light on its travel on the line of
sight. Such mechanism was considered by Gnedin 1994 and by Gnedin
and Krasnikov 1992 and it consists in photon-pseudoscalar mixing
within a magnetic field. Mixing process produces light
polarization in an external magnetic field. The coupling between
electromagnetic and pseudoscalar fields affects the polarization
properties of the electromagnetic waves as they propagate through
a magnetic field. The value of linear polarization due to magnetic
conversion of photons into pseudoscalar bosons is derived by the
expression:

\begin{equation}
P_l = A \sin^2{(\frac{1}{2} g_{\alpha \gamma} B_{\bot} l)}
\label{eq2}
\end{equation}

\noindent where $B_{\bot} = B\sin{\theta}$ is the perpendicular
component of a magnetic field, $\theta$ is the angle between the
photon propagation and magnetic field directions, l is the
characteristic dimension of a magnetic field, $g_{\alpha \gamma}$
is the photon-pseudoscalar coupling constant. The alignment effect
requires existence of regular magnetic field on the cosmological
scales $\sim 1$ Gpc with the magnetic field strength at the level
of $\sim 1 \div 10$ nG. There is now evidence of existence of such
huge magnetic fields on cosmological scales (Blasi and Olinto,
1999, Dar and De Rujula, 2005). This evidence was obtained via
observations of Faraday rotation of light from distant QSOs.

\section{Measurement of the spectrum of linear polarized
radiation\\ from magnetosphere of compact object - a new method to
determine the magnetic field}

The existing magnetic field of the astrophysical object is the
additional factor in optical anisotropy of a surrounding this
object plasma. The radiation of such object acquires linear
polarization as a result of scattering on electrons. The scattered
radiation undergoes Faraday rotation by propagation in magnetized
plasma. The angle of Faraday rotation $\psi$ is determined by the
expression (1). The integral linear polarization of light occurs
even for a spherically symmetric density distribution of electrons
if the magnetic field has no axial symmetry coinciding with the
line of sight. Fig.4 demonstrates action of this effect. The
radiation only scattered in equatorial volumes is polarized
because Faraday rotation angle is negligible for radiation
direction across magnetic force lines.

Another important effect is the polarization of radiation
scattered from an accretion disk that also undergoes to Faraday
rotation and Faraday depolarization. Photons escape the optically
thick disk basically from the surface layer with $\tau \sim 1$. If
the Faraday rotation angle corresponding to this optical length
becomes greater than unity, then the emerging radiation will be
depolarized as a result of the summarizing radiation fluxes with
different angles of Faraday's rotation. Only for directions that
are perpendicular to the vertical magnetic field the Faraday
rotation angle is too small to yield depolarization effect.
Certainly, the diffusion of radiation in the inner part
depolarizes it even without magnetic field because of multiple
scattering of photons. The Faraday rotation only increases the
depolarization process. It means that the polarization of outgoing
radiation acquires the peak-like angular dependence with its
maximum for the direction perpendicular to magnetic field. The
sharpness of the peak increases with increasing magnetic field
strength. The main region of allowed angles appears to be $\sim
1/\delta$, where the parameter depolarization:

\begin{equation}
\delta = 0.8 \left(\frac{\lambda}{1\mu m}\right)^2
\left(\frac{B}{1 G}\right)
\label{eq3}
\end{equation}

Fig.5 shows the dependence of the polarization degree of outgoing
radiation from magnetized optically thick plasma disk. The maximum
value of polarization is equal to 9.14\% instead of the classical
value 11.7\% corresponding to multiple electron scattering without
magnetic field. The results of our calculations (see Gnedin et
al.2005) polarization for the various models of optically thick
accretion disk with vertically averaged magnetic field are
presented by the Table 1.

\begin{table}
\caption[]{Degree of polarization for the various models of an
accretion disk with vertically averaged magnetic field.}
\begin{flushleft}
\begin{tabular}{|l|l|l|}
\hline

{\bf Model}& $B_{eq}(R)$& $P_l(\lambda)$ \\

\hline

{\small Accretion disk with ion}& & \\
{\small supported flows}&
$\sim R^{-5/4}$& $\sim \lambda^{-1/3}$ \\

\hline

{\small Sunayev-Shakura disk (a)}& & \\
{\small $P_r \gg P_g$}&
$\sim R^{-3/4}$& $\sim \lambda^{-1}$ \\

\hline

{\small Sunayev-Shakura disk (b)}& & \\
{\small $P_g \gg P_r$}&
$\sim R^{-9/8}$& $\sim \lambda^{-1/2}$ \\

\hline

{\small Sunayev-Shakura disk (c)}& & \\
{\small $P_g \gg P_r$}&
$\sim R^{-21/16}$& $\sim \lambda^{-1/4}$ \\

\hline

{\small Hot accretion disk with}& & \\
{\small plasma viscosity}&
$\sim R^{-15/28}$& $\sim \lambda^{-9/7}$ \\

\hline

{\small Payne-Eardley disk}& & \\
{\small $P = P_g$, $\alpha =
1$}& $\sim R^{-21/8}$& $\sim \lambda^{-1/8}$ \\

\hline

{\small Magnetic accretion-jet ejection}& & \\ {\small disk
without equipartition}& $\sim R^{-5/2}$& $\sim \lambda^{4/3}$ \\

\hline

{\small Accretion disk with non-zero}& & \\ {\small torque on
inner edge}& $\sim R^{-15/16}$& $\sim \lambda^{-1}$ \\

\hline

{\small Disk with reprocessing}& $\sim R^{-7/4}$& $\sim
\lambda^{-1/8}$ \\

\hline
\end{tabular}
\end{flushleft}
\end{table}
AGNs and QSOs are characterized with strong outflows from
accretion disks. Punsly (2001) developed the theory that a wind of
magnetized plasma from the nearest vicinity of supermassive black
hole can be driven by the interaction of the black hole
gravitational field and a plasma filled magnetosphere. He
investigated the fundamental process that allows a rotating black
hole to power a magnetized wind. The basic Punsly's idea is that
the poloidal magnetic field is generated in the region between the
inner radius of the accretion disk and the horizon (ergosphere).
Our basic idea is that cooperative action of the wind outflow and
rotation of a black hole can transfer the poloidal magnetic field
in the magnetic field of Parker type:

\[
B_r = B \left(\frac{R_S}{r}\right)^2 \cos{\theta}
\]
\[
B_{\varphi} = B \frac{k a}{M} \frac{1}{1 -
a/M}\left(\frac{R_S}{r}\right)\cos{\theta}\sin{\theta} \equiv C
\left(\frac{R_S}{r}\right) \cos{\theta} \sin{\theta}
\]
\begin{equation}
B_{\theta} = 0
\label{eq4}
\end{equation}

\noindent Here $B$ is the polar magnetic field at the inner radius
$R_S$ of the accretion outflow, $\theta$ is the angle between the
dipole axis and the radius-vector, $a/M$ is the well-known
dimensionless black hole angular momentum parameter, k is the
specific physical parameter $\sim 1$ of a black hole ergosphere.
The results of our calculation of polarization are presented at
Fig.6. This Fig. reveals that if the toroidal component is much
less that and radial one the scattered radiation is nearby
depolarized. With increasing toroidal component, i.e. with
reaching the value $a/M \sim 1$ the polarization spectrum removes
in more hard wavelength region, the width of spectral dependence
being narrow. In principle, the increase of the radial component
makes also shift of the maximum polarization into hard energy, but
in this case much more stronger magnetic field strength is
required. The real conclusion is that more faster rotating black
holes will have higher level polarization compared to slowly
rotating black holes. It means that the level of polarization and
its spectral distribution is acceptable test for distinguishing
between Kerr and Schwarzschild black holes.

The plasma outflow from the magnetosphere and the inner part of
the accretion disk gives rise to an envelope. The radiation
scattered in the envelope acquires the linear polarization. The
region of the formation of broad emission lines observed in AGNs
is a case of such envelope. In the absence of magnetic field the
scattered radiation has to be non -polarized provided the
scattering matter in the spherically shape envelope is symmetric
relative to the line of sight. If the magnetic field is not
symmetric relative to this line, the outgoing radiation (singly or
multiply scattered) has to undergo influence of the Faraday's
rotation effect  and become to be polarized. The integral
polarization of radiation is equal to zero if the dipole magnetic
moment M is coaxial with the line of sight or in the case of $B =
0$. The results of numerical calculations of polarization are
presented in Fig.7 as function of dimensionless parameter
$\delta$, where $B = M/a^3$ is the dipole magnetic field on the
surface of radiation source. This parameter changes from 0 up to
100. The peak value of the polarization occurs when $M$ is
perpendicular to the line of sight.

\section{Magnetic fields of the extragalactic radio sources:
exploration of the cosmological models}

We demonstrate the principal possibility to explore the various
cosmological models through measurements of the magnetic field
magnitudes of the compact radio galaxies. The magnetic field
strengths of the compact extragalactic radio structures can be
estimated from the synchrotron spectra of radio sources if one
takes into account the synchrotron self-absorption process. In
this case the most important physical value is the observed
synchrotron self-absorption frequency $\nu_p$ at that the optical
thickness respect to the synchrotron self-absorption becomes equal
to unite. Then the expression for the magnetic field strength $B$
is determined by Slysh,1963, and Hirotani, 2005:

\begin{equation}
B = 10^{-5} b(\alpha) \left(\frac{\nu_p}{1 GHz}\right)^5
\left(\frac{\theta}{mas}\right)^4 \left(\frac{1 Jy}{S_p}\right)^2
\frac{\delta}{1 + z}
\label{eq5}
\end{equation}

\noindent Here $S_p$ is the flux density at the frequency $\nu_p$,
$\delta$ is the boosting factor, the coefficient $b(\alpha)$
depends on the index of synchrotron radiation. Its value lies in
the region $b = 2 \div 3$ for the wide range of values $\alpha$.

This formula was successively used by Artukh and Chernikov, 2001,
Tyul'bashev and Chernikov, 2001, for determination of magnetic
fields of concrete extragalactic radio sources.

Our main idea is to use this formula for exploration various
cosmological models using in this formula the well-known
expression for angular diameter $\theta$ of the radio source:

\begin{equation}
\theta = \frac{l (1 + z)^2}{D_L (z)}
\label{eq6}
\end{equation}

\noindent where $D_L$ is, so-called, the photometric distance.
This distance depends on the real cosmological model. For the
commonly accepted the dark matter (DM) and dark energy (DE) model
the expression for the photometric distance takes the form:

\begin{equation}
D_L (z) = \frac{c}{H_0} (1 + z)\int_0^z{\frac{dx}{H(x)/H_0}}
\label{eq7}
\end{equation}

\noindent Where the Hubble parameter is:

\begin{equation}
H(z) = H_0 \left[\Omega_m (1+z)^3 + \Omega_{\Lambda}
f(z)\right]^{1/2}
\label{eq8}
\end{equation}

$\Omega_m$ and $\Omega_{\Lambda}$ are the ratios of the modern
total matter density and dark energy density to the critical
density, correspondingly, and what is more $\Omega_m +
\Omega_{\Lambda} = 1$. The function $f(z)$ depends on the equation
of state of dark energy $p(z) = w(z)\rho(z)$:

\begin{equation}
f(z) = \exp{\left[3 \int_0^z{\frac{1 + W(z)}{1 + z} dz}\right]}
\label{eq9}
\end{equation}

One can see that the magnitude of magnetic field depends very
strongly on the photometric distance and therefore on the real
cosmological model. The relation (9) means also that the value
magnetic field in the radio luminosity region depends essentially
on the form of the equation state of dark energy. Such dependence
appears to be more strongly that it looks in the classical methods
of SNIa and cosmic microwave background anisotropy (WMAP).

We demonstrate the power of this method on the example of the
sample of the Gigahertzs Peaked Spectrum sources (GPS) that
presents the class of intrinsically compact objects (linear size
at parsec scale). This sample defined on the basis of their
spectral properties (see Tinti et al., 2004): the overall shape is
convex with turnover frequencies of a few GHz and the spectrum at
high frequencies is steep. The best fitting physical mechanism is
the synchrotron emission of the relativistic electrons with the
synchrotron self-absorption (SSA).

The key problem is the determination of the photometric distance
which is really dependent on the basic parameters DM and DE. We
are considering the some examples of the cosmological models when
$\Omega_m + \Omega_{\Lambda} = 1$ and $P_{DE} = W \rho_{DE}$,
where $W > < 0$ for different cosmological models and then:

\begin{equation}
H(z) = H_0 \left[\Omega_m (1 + z)^3 + \Omega_{\Lambda} (1 + z)^{3
(1 + W)}\right]^{1/2} \label{eq10}
\end{equation}

For popular LCDM (Lambda-Cold Dark Matter) model it follows $W =
-1$, for model with noninteracting cosmic strings $W = - 1/3$. In
the Universe with domain walls the DE equation of state requires
$W = - 2/3$. At last, in the so-called phantom model it appears
that $W < - 1$. Our calculations of the Eq.(5) show that the
magnitude of the magnetic field of one definite radio source can
differ at 2 - 3 times for various cosmological models mentioned
here.

\begin{table}
\caption[]{Magnetic field strengths for GPS source 1424+2256 (z =
3.626) in various cosmological models.}
\begin{flushleft}
\begin{tabular}{|l|l|}
\hline

{\bf Model}& {\bf B(mG)} \\

\hline

LA (3)& 0.152 \\

\hline

P2 (4)& 0.129 \\

\hline

Linear (5)& 0.245 \\

\hline

CA (7)& 0.034 \\

\hline

Quiess (8)& 0.079 \\

\hline

MCG (9)& 0.085 \\

\hline

Brane 2 (10)& 0.083 \\

\hline

LCDM& 0.080 \\

\hline
\end{tabular}
\end{flushleft}
\end{table}
We present in the Table 2 the results of calculations of magnetic
fields of the GPS source ($z = 3.626$) for various cosmological
models. The names of the models presented in this Table are: LA
corresponds to the following expression for DE coefficient $W(z) =
W_0 + W_1 z / (1 + z)$ and $W_0 = - 1.40$, $W_1 = 1.66$ (Linder,
2003). Linear means $W(z) = W_0 + W_1 z$ with the same values of
$W$'s. P2 is a quadratic polynomial of $H(z)$ (Alam et al.,2004).
P3 is a cubic polynomial $H(z)$. CA is the generalized cardassian
cosmology (Freese, 2005). Quiess means the DE with constant
equation of state $W$. MCG is a modified form (Chimento et al.
2004) of the Chaplygin gas (Fabris et al. 2001). Brane 2 means
brane world cosmology (Sahni and Shtanov, 2003, Sahni and Alam,
2002).

One can see that there is a real difference in magnitudes of
magnetic fields for various cosmological models. It means that
measurements of magnetic fields of active galaxy nuclei and
quasars allow us to choose the real cosmological model from
numerous theoretical models of evolution of our Universe.

The basic difference between the cosmological models can be
formulated by the following way. Magnetic fields of AGNs in the
phantom cosmological model with $W < -1$ is less than in the
cosmological models with $W \geq -1$, i.e. $B(W < -1) < B(W =
-1)$. We reveal the following relation between magnitudes of
magnetic fields for GPS 1424+2256 ($z = 3$) for Standard Cold
Matter (SCDM) and commonly accepted LCDM models:

\begin{equation}
B_{SCDM} / B_{LCDM} = 7
\label{eq11}
\end{equation}

Magnetic field is greater for brane world model and Chaplygin gas
model compare to classical LCDM model. For curvature theory of
gravity the magnetic field also is strongly greater than for LCDM
model, and what is more $B_{curv} / B_{LCDM}$ can reach $\sim
100$. Our calculations show also that magnetic fields for the same
source in positive curvature universe and in negative curvature
universe can differ each from other, at least, two times.

It is interesting to consider the situation with dynamical
behavior of dark energy. There is an evidence to show that the
dark energy might evolve from $W > -1$ in the past to $W < -1$
today and cross $W = -1$ in the intermediate redshift (see, for
example, Huang and Guo, 2005). The basic question is what explicit
value of transition redshift. There is some evidence from SNIa,
that transition from decelerating to accelerating  takes place at
$z \sim 0.6$, but Huang and Guo, 2005, found that this transition
takes at $z \sim 1.7$. We calculate the magnetic field strength
for the radio source GPS 1424+2256 ($z = 3.626$) with and without
transition from decelerating to accelerating. If the transition
takes at $z \sim 0.6$, the ratio $B_{tr} / B_{LCDM} = 5.4$. If the
transition takes place at $z = 1.7$ then $B_{tr} / B_{LCDM} =
2.7$. This result means that it is possible to distinguish two
various transitions: $z \sim 0.6$ and $z \sim 1.7$ via
measurements of magnetic field. For $z = 3.626$ the ratio $B(0.6)
/ B(1.7) = 2.27$. Thus the future measurements of magnetic fields
of active galaxy nuclei and quasars will be able make decisive
contribution to solution of the problem of evolution of the
Universe.
\begin{table}
\caption[]{Magnetic field strengths for a number of GPS sources in
various cosmological models (with different deceleration
parameters).}{\small
\begin{flushleft}
\begin{tabular}{|l|l|l|l|l|l|}
\hline

{\bf Object}& & $q_0 = 0$& $q_0 = 0.5$& WDE =& WDE =\\ & & & & -1&
-1 + 2z \\

\hline

1840+3900& $B(l_0 / l)^4$(G)& 0.0015& 0.0195& 0.0035& 0.0286 \\

 & $B / B_0$& 1.000& 13.120& 2.381& 19.184 \\

\hline

0646+4451& $B(l_0 / l)^4$(G)& 0.0006& 0.0095& 0.0016& 0.0158 \\

 & $B / B_0$& 1.000& 15.758& 2.740& 26.342 \\

\hline

1424+2256& $B(l_0 / l)^4$(G)& 0.0000& 0.0005& 0.0001& 0.0009 \\

 & $B / B_0$& 1.000& 18.023& 3.113& 32.993 \\

\hline

1616+0459& $B(l_0 / l)^4$(G)& 0.0000& 0.0004& 0.0001& 0.0006 \\

 & $B / B_0$& 1.000& 13.974& 2.463& 21.427 \\

\hline

1645+6330& $B(l_0 / l)^4$(G)& 0.0186& 0.1521& 0.0299& 0.1534 \\

 & $B / B_0$& 1.000& 8.163& 1.603& 8.235 \\

\hline

1850+2825& $B(l_0 / l)^4$(G)& 0.0003& 0.0029& 0.0006& 0.0033 \\

 & $B / B_0$& 1.000& 9.255& 1.777& 10.537\\

\hline
\end{tabular}
\end{flushleft}
}
\end{table}
In the Table 3 the results of calculations of magnetic field
strength for the  number of the GPS  sources are presented for the
cosmological models with different deceleration parameter $q_0$.
The scaling linear dimension of radio region $l_0$ is chosen as
$l_0 = 2$ pc. Our results allows to determine the magnitude of
magnetic field in the close vicinity of the supermassive black
hole horizon. If one suggests the conservation of the magnetic
flux one can estimate this magnitude. The data of the Table 3 show
that this magnitude can reach the value $B_H \sim 10^8 \div 10^9$G
at the horizon of a supermassive black hole.

\section{Conclusions}

The alignment of the polarization vectors of AGNs and QSOs
discovered by Hutsemekers and their colleagues may be explained as
the effect of conversion of photons into pseudoscalar bosons  in
the intergalactic magnetic field. Polarimetric observations  in
optical range and radio spectral observations of AGNs and QSOs
allow us to measure the magnetic field magnitudes. It is
interesting that the magnetic field strength  determined from
radio spectrum in the model of synchrotron self-absorption
mechanism appears to be strongly dependent on the cosmological
model. It means that magnetic measurements are presented a new
test for exploration of cosmological models and for determination
of the basic parameters of dark matter and dark energy.

\section*{Acknowledgements}

This work was supported by the RFBR grant 03-02-17223, the Program
of the Presidium of RAS "Nonstationary Phenomena in Astronomy",
the Program of the Department of Physical Sciences of RAS "The
Extended Structure...", and by the Program of Russian Education
and Science Department. One of the authors (M.Yu. Piotrovich) is
grateful to "Russian Science Support Foundation".

\bigskip

\section*{Figures captions}

{\it (See files: 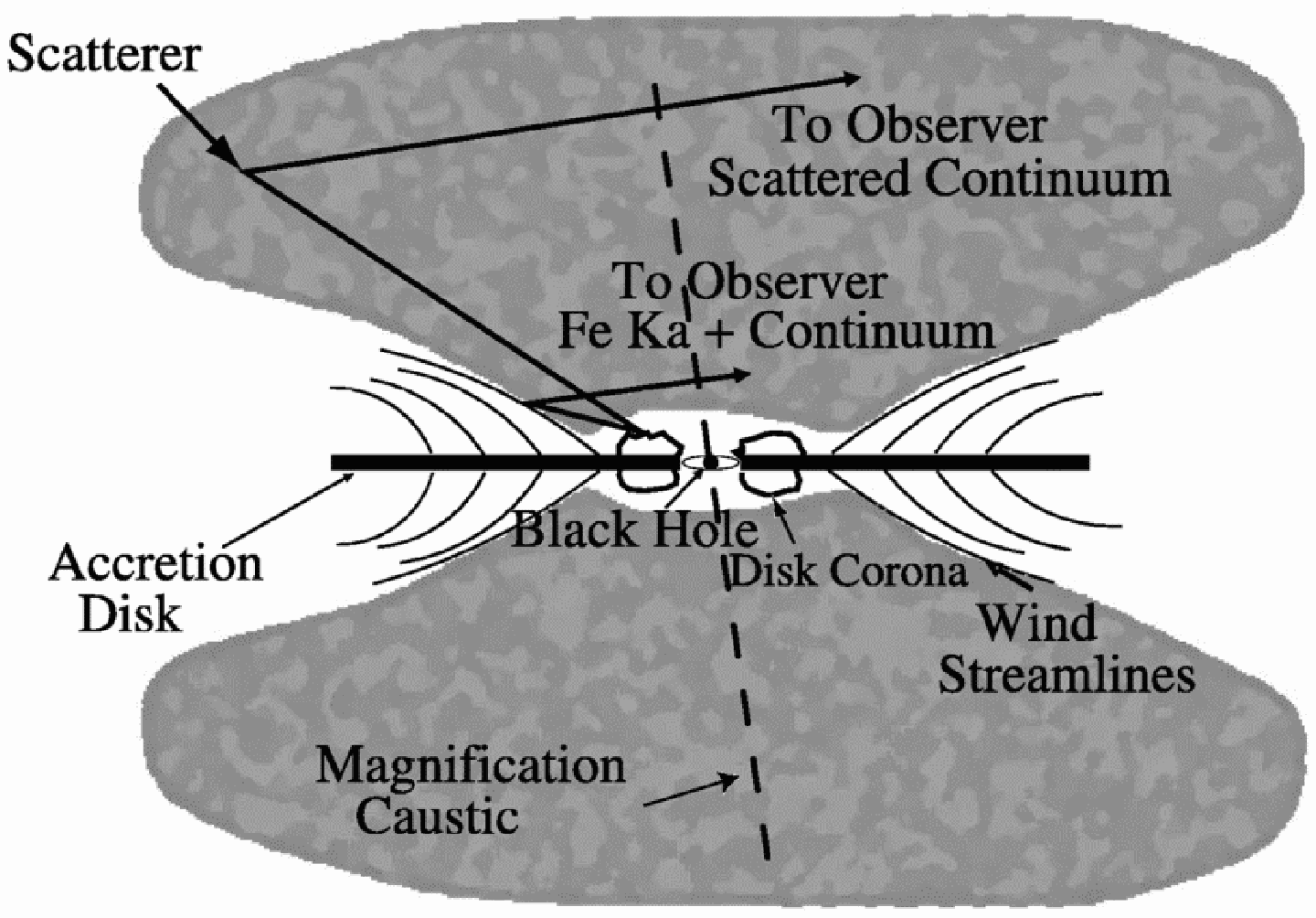, 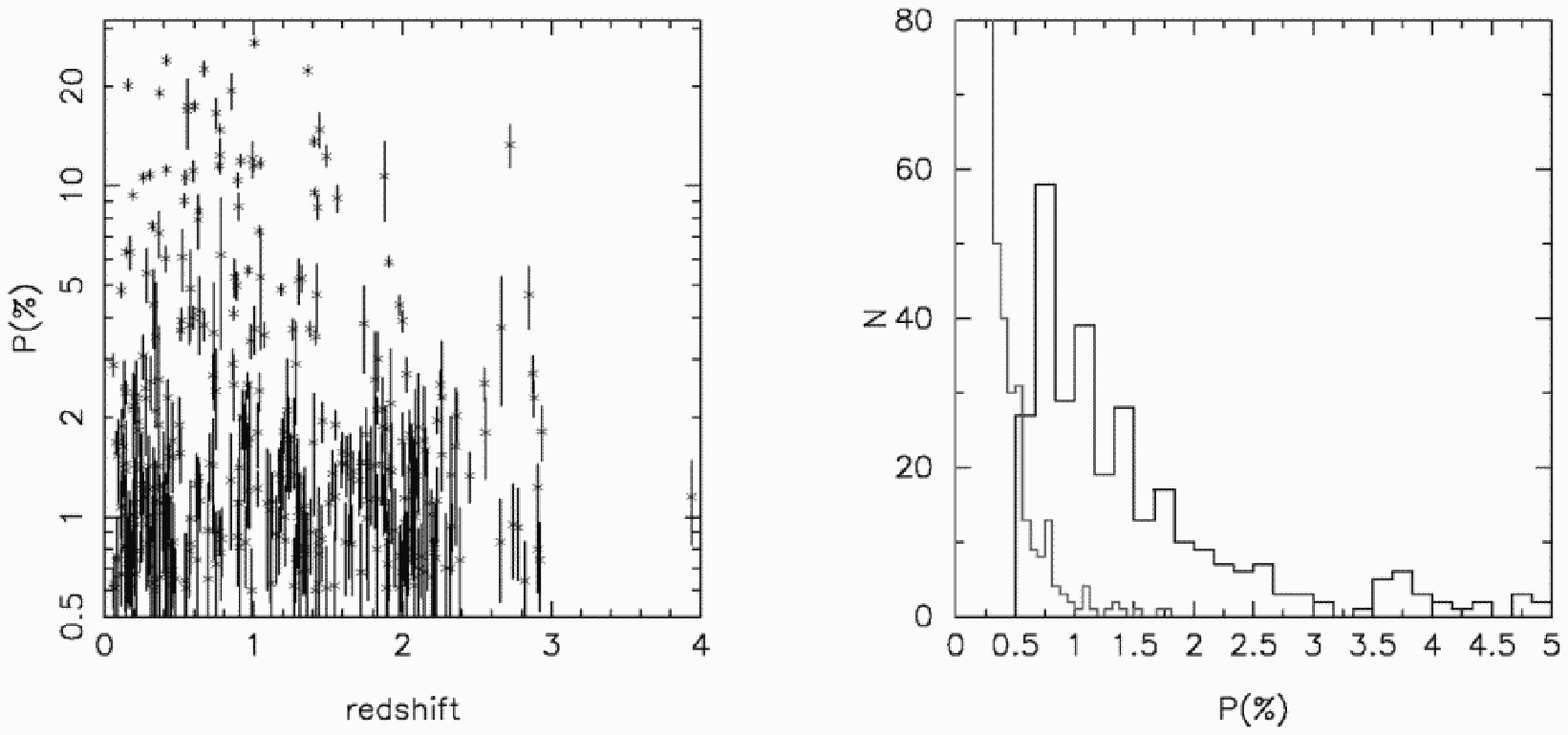, 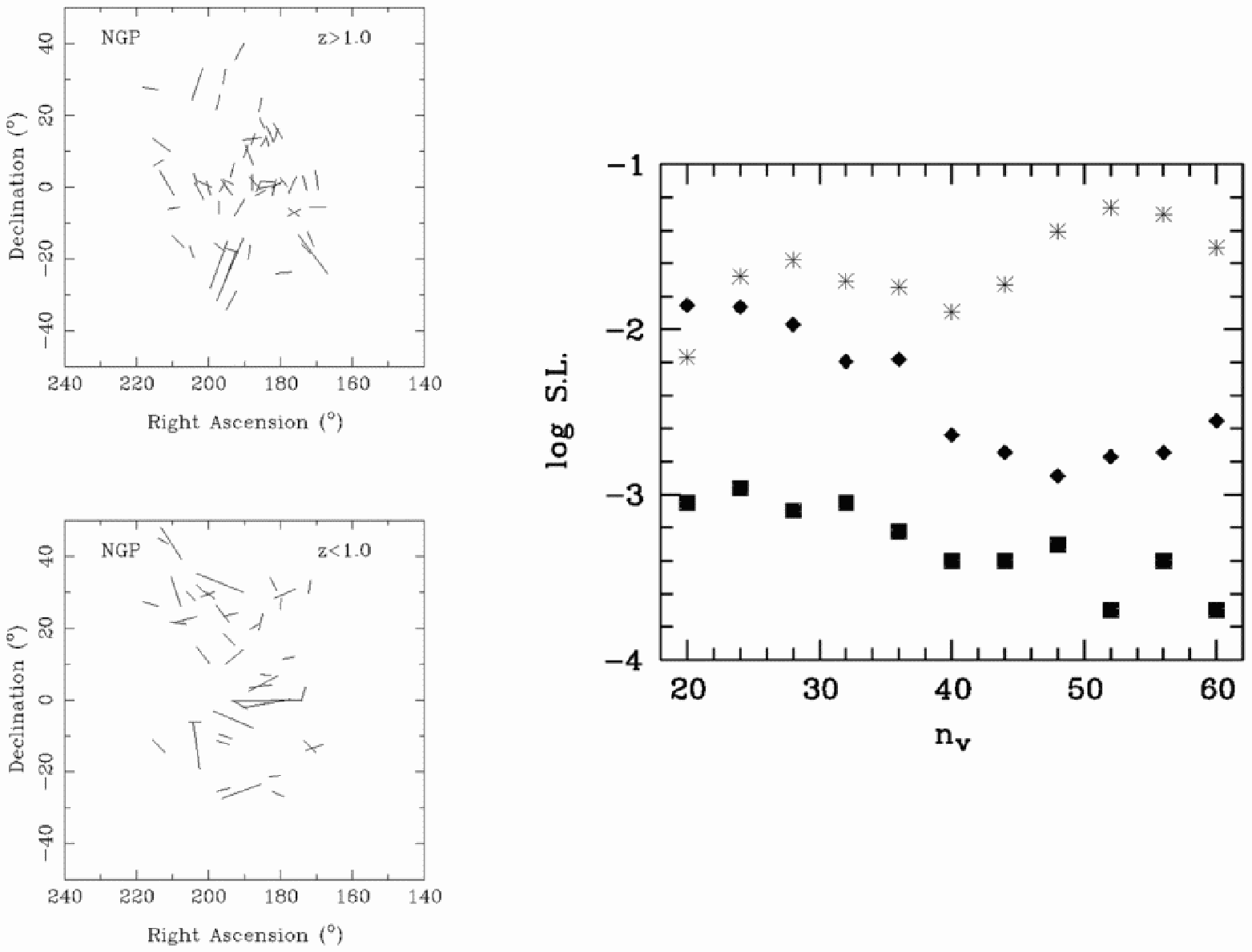, 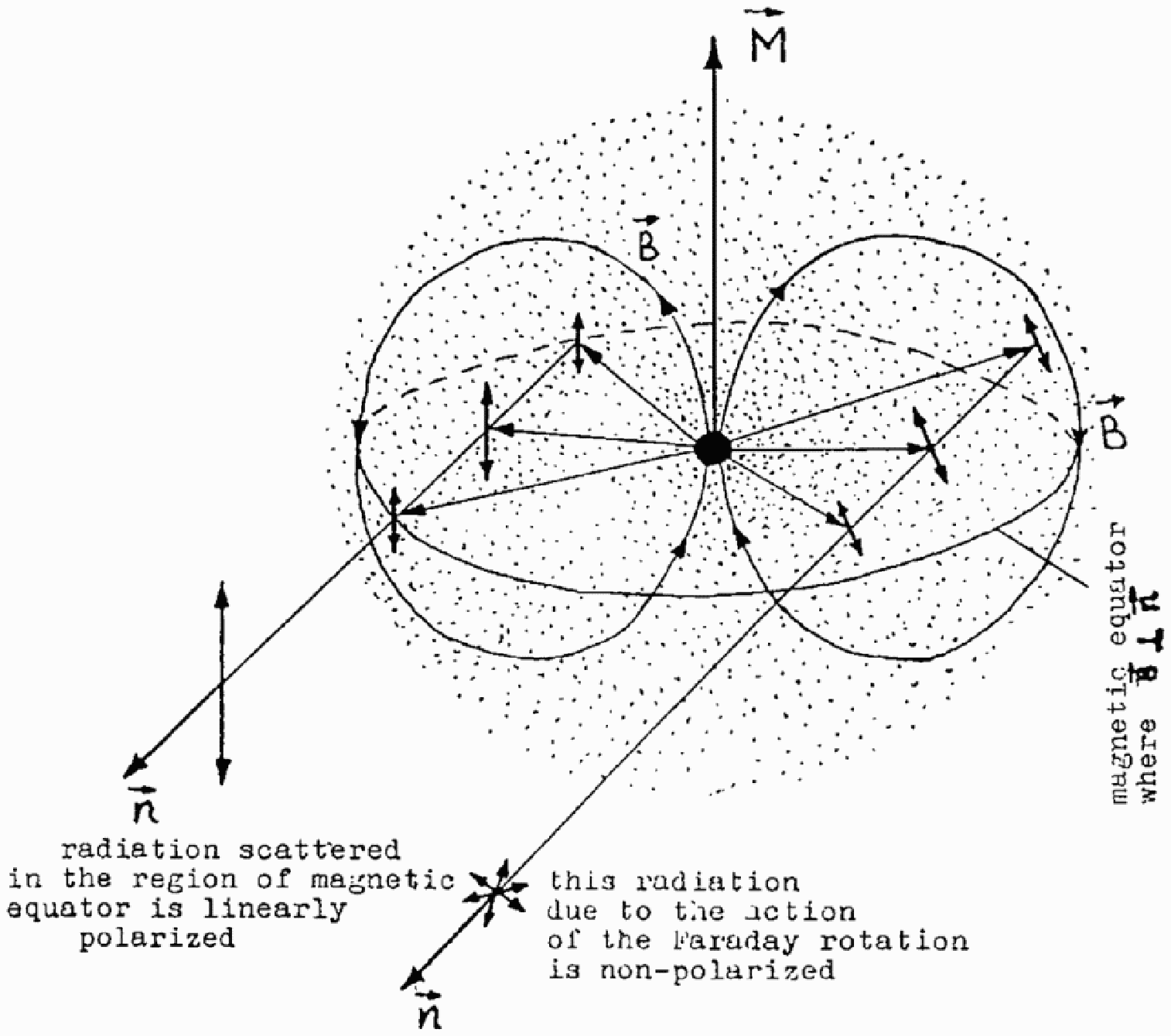, 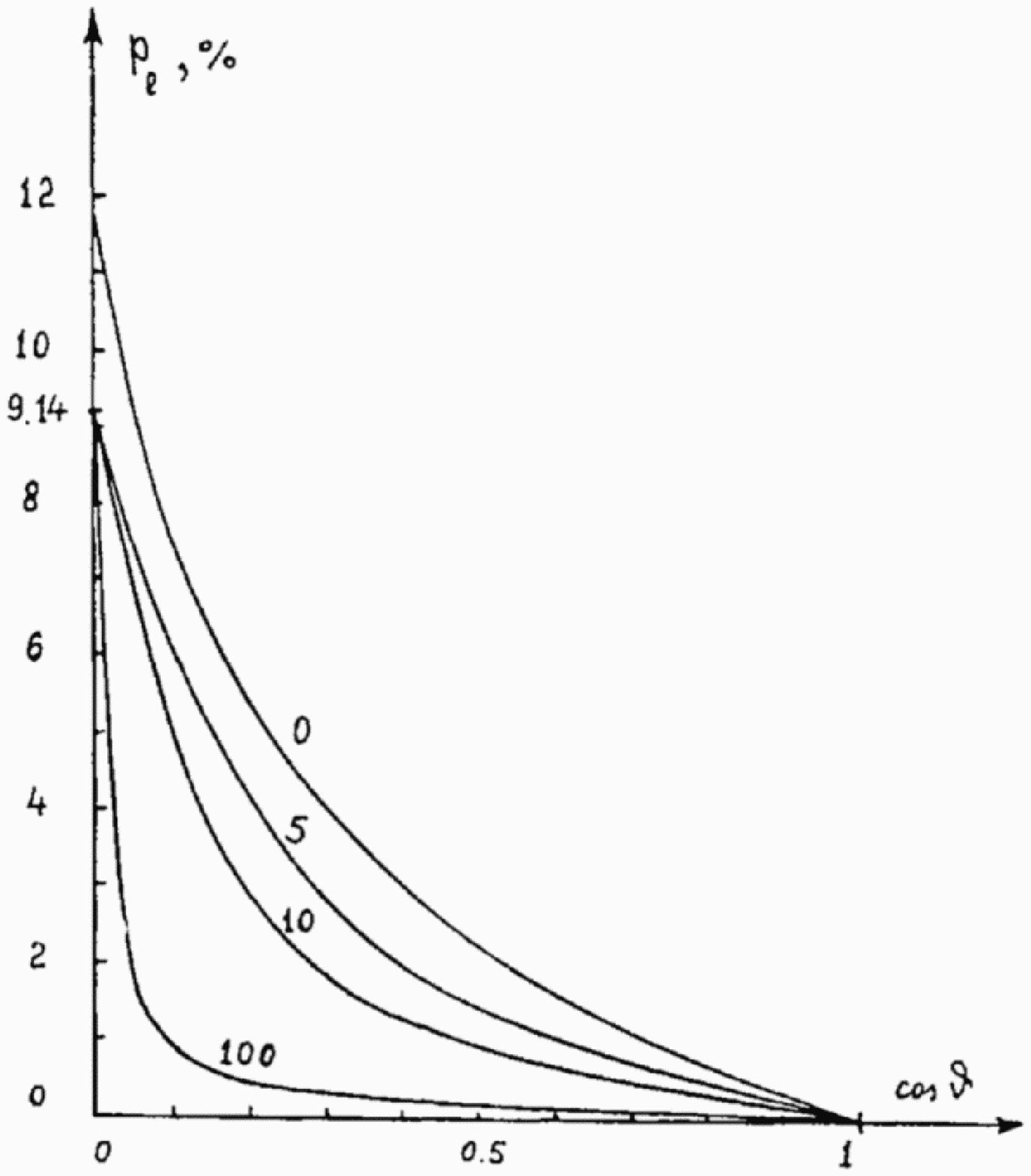,
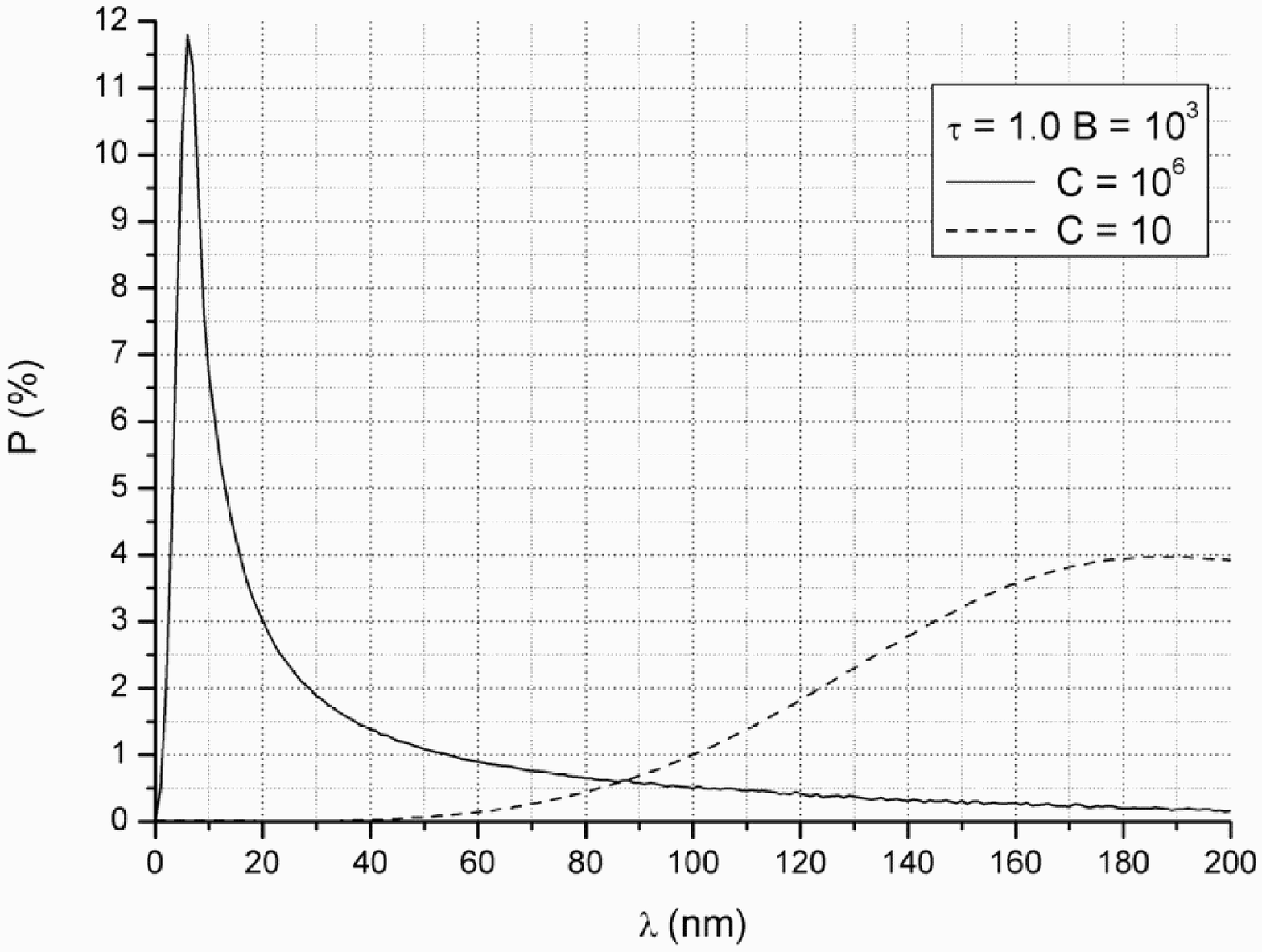, 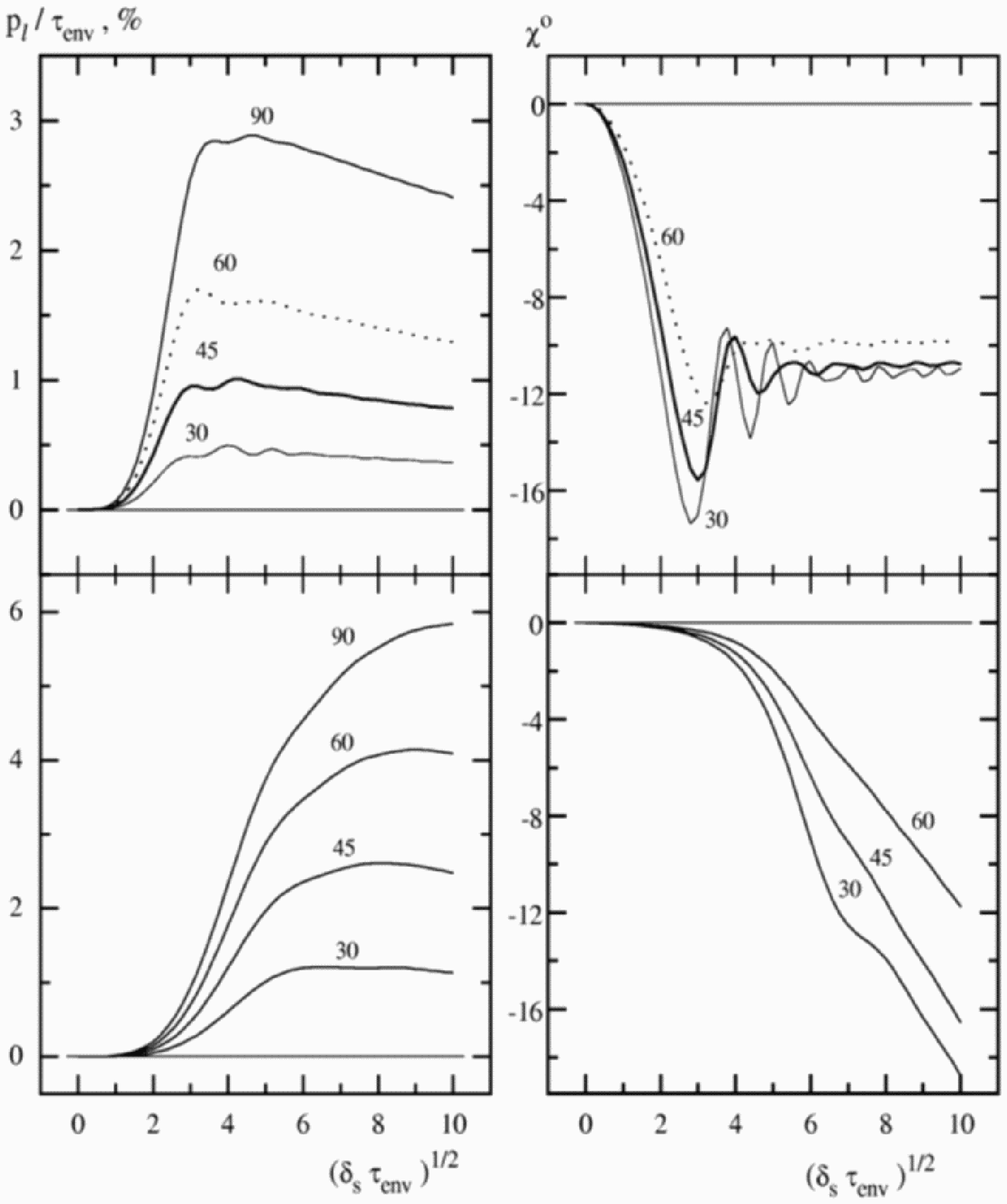.)}

\bigskip

{\bf Fig.1.} Structure of the AGN nearest region: accretion disk,
corona and outflow wind lines.

\medskip

{\bf Fig.2.} The results of polarization measurements of the
sample chosen by Cabanac et al., 2005.

\medskip

{\bf Fig.3.} The confidence level of electric vectors alignment
for QSOs (from Cabanac et al.,2005).

\medskip

{\bf Fig.4.} Mechanism of production of polarized radiation in the
spherically symmetric scattering envelope with a dipole magnetic
field.

\medskip

{\bf Fig.5.} Dependence of the polarization of radiation from the
optically thick disk on the angle between the line of sight and
the normal to the disk.

\medskip

{\bf Fig.6.} Spectrum of polarized radiation scattered in the
outflow from a rotating supermassive black hole. The magnetic
field is suggested Parker-type structure.

\medskip

{\bf Fig.7.} Results of calculation of polarization of radiation
scattered in the spherical outflow with a dipole magnetic field
from a supermassive black hole.


\begin{thebibliography}{99}
\bibitem{1}
Gnedin Yu.N., Silant'ev N.A. {\it in} ``Basic Mechanisms of Light
Polarization in Cosmic Media'', Amsterdam, Hartwood Academic
Publ., p.1 - 49, 1997.
\bibitem{2}
Cabanac R., Hutsemekers D., Sluse D., Lamy H. 2005, {\it ASP Conf.
Series}, in press, astro-ph/0501043, 2005.
\bibitem{3}
Veron-Cetty M.P., Veron P., {\it Astron. Astrophys.}, {\bf 374},
92 (2001).
\bibitem{4}
Hutsemekers D., Lamy H., 2001, {\it Astron. Astrophys.}, {\bf
361}, 381 (2001).
\bibitem{5}
Hutsemekers D., Cabanac R., Lamy H., Sluse D., astro-ph/0507274
(2005).
\bibitem{6}
Gnedin Yu.N., {\it Astron. Astrophys. Transactions}, {\bf 5}, 163
(1994).
\bibitem{7}
Gnedin Yu.N., Krasnikov S.V., {\it Sov. Phys.(JETP)}, {\bf 75},
933 (1992).
\bibitem{8}
Blasi P., Olinto A., {\it Phys. Rev.}, {\bf D59}, 023001 (1999).
\bibitem{9}
Dar A., De Rujula A., astro-ph/0504480, (2005).
\bibitem{10}
Gnedin Yu.N., Silant'ev N.A., Shternin P.S., {\it Astron. Rep.},
in press, astro-ph/0503121, (2005).
\bibitem{11}
Punsly B., ``Black Hole Gravitohydromagnetics'', Springer, 2001.
\bibitem{12}
Slysh V.I., {\it Nature}, {\bf 199}, 682 (1963).
\bibitem{13}
Hirotani K., {\it ApJ}, {\bf 619}, 73 (2005).
\bibitem{14}
Artyukh V.S., Chernikov P.A., {\it Astron. Zh.}, {\bf 78}, 20
(2001).
\bibitem{15}
Tyul'bashev S.A., Chernikov P.A., {\it Astron. Astrophys.}, {\bf
373}, 381 (2001).
\bibitem{16}
Tinti S., Dallacassa D., De Zotti G. et al., astro-ph/0410663
(2004).
\bibitem{17}
Linder E.V., {\it Phys. Rev.}, {\bf D68}, 083503 (2003).
\bibitem{18}
Alam U., Sahni V., Starobinsky A.A., {\it JCAP}, {\bf 0406}, 008
(2004).
\bibitem{19}
Freese K., astro-ph/0501675 (2005).
\bibitem{20}
Chimento L.P., {\it Phys. Rev.}, {\bf D69}, 123517 (2004).
\bibitem{21}
Fabris J.C., Goncalves S.V.B., de Souza P.E., {\it Gen. Rel.
Grav.}, {\bf 34}, 53 (2002).
\bibitem{22}
Sahni V., Shtanov Y., {\it JCAP}, {\bf 0311}, 014 (2003).
\bibitem{23}
Sahni V., astro-ph/0502032 (2005).
\bibitem{24}
Huang C.-G., Guo H.-Y., astro-ph/0508181 (2005).
\end{thebibliography}
\end{document}